\documentclass[prl,twocolumn,superscriptaddress,showpacs]{revtex4}
\usepackage{times}
\usepackage{amsmath}
\usepackage{amsfonts}
\usepackage{amssymb}
\usepackage{graphicx}
\usepackage{pst-all}

\begin{document}

\title{Measurement-Only Topological Quantum Computation}
\author{Parsa Bonderson}
\affiliation{Microsoft Research, Station Q, Elings Hall, University of California, Santa Barbara, CA 93106}
\author{Michael Freedman}
\affiliation{Microsoft Research, Station Q, Elings Hall, University of California, Santa Barbara, CA 93106}
\author{Chetan Nayak}
\affiliation{Microsoft Research, Station Q, Elings Hall, University of California, Santa Barbara, CA 93106}
\affiliation{Department of Physics, University of California, Santa Barbara, CA 93106}
\date{\today}

\begin{abstract}
We remove the need to physically transport computational anyons around each other from the implementation of computational gates in topological quantum computing. By using an anyonic analog of quantum state teleportation, we show how the braiding transformations used to generate computational gates may be produced through a series of topological charge measurements.
\end{abstract}

\maketitle


Topological quantum computation (TQC) is an approach to quantum computing that derives fault-tolerance from the intrinsically error-protected Hilbert spaces provided by the non-local degrees of freedom of non-Abelian anyons~\cite{Kitaev03,Preskill98,Freedman02a,Freedman02b}. To be computationally universal, the anyon model describing a topologically ordered system must be intricate enough to permit operations capable of densely populating the computational Hilbert space. At its conception, the primitives envisioned as necessary for implementing TQC were:

1. Creation of the appropriate non-Abelian anyons, which will encode topologically protected qubits in their non-local, mutual multi-dimensional state space.

2. Measurement of collective topological charge of anyons, for qubit initialization and readout.

3. Adiabatic transportation of computational anyons around each other, to produce braiding transformations that implement the desired computational gates.

In an effort to simplify, or at least better understand the TQC construct and what is essential to its architecture, we reconsider the need for physically braiding computational anyons. We demonstrate that it is not a necessary primitive by replacing it with an adaptive series of non-demolitional topological charge measurements. Naturally, this ``measurement-only'' approach to TQC draws some analogy with other measurement-only approaches of quantum computing~\cite{Raussendorf01,Nielsen03,Raussendorf06,Bravyi07}, but has the advantage of not expending entanglement resources, thus allowing for computations of indefinite length (for fixed resource quantities).

In this letter, we only consider orthogonal, projective measurements~\cite{vonNeumann55} of topological charge, for which the probability and state transformation for outcome $c$ is given by
\begin{eqnarray}
\text{Prob}\left(c\right) &=& \left\langle \Psi \right| \Pi_{c} \left| \Psi \right\rangle \\
\label{eq:postmeasstate}
\left| \Psi \right\rangle &\mapsto& \frac{\Pi_{c} \left| \Psi \right\rangle}{\sqrt{\left\langle \Psi \right| \Pi_{c} \left| \Psi \right\rangle}}
.
\end{eqnarray}
Such measurements potentially include Wilson loop measurements in lattice models, energy splitting measurements in fractional quantum Hall (FQH) and possibly other systems, and (the asymptotic limit of) interferometry measurements when the measured charge $c$ is Abelian. Though these may involve or be related to the motion of some anyons, the measured anyons are not moved (at least not around each other).

We use a diagrammatic representation of anyonic states and operators acting on them, as described by an anyon model (for definitions, including normalizations, see e.g.~\cite{Bonderson07b,Bonderson07c}). Through a set of combinatorial rules, these diagrams encode the purely topological properties of anyons, independent of any particular physical representation. Using this formalism, we show how projective measurement of the topological charge of pairs of anyons enables quantum state teleportation. We then show that repeated applications of teleportation can have the same effect (up to an overall phase) as a braiding exchange of two anyons. Multiple applications of this protocol then allow any desired braid to be mimicked without ever having to move the anyons.

Within the isotopy invariant diagrammatic formalism, the projector onto definite charge $c$ of anyons $1$ and $2$ (numbered from left to right) of definite charges $a$ and $b$ is~\footnote{We only consider anyon models with no fusion multiplicities (i.e. the fusion coefficients $N_{ab}^{c}=0,1$) in this letter, but the analysis may be generalized.}
\begin{equation}
\Pi_{c}^{\left(12\right)} = \left| a,b;c \right\rangle \left\langle a,b;c\right|
= \sqrt{\frac{d_{c}}{d_{a}d_{b}}} \quad
 \pspicture[0.4138](0,-0.45)(1.5,1)
 \small
  \psset{linewidth=0.9pt,linecolor=black,arrowscale=1.5,arrowinset=0.15}
  \psline{->}(0.7,0)(0.7,0.45)
  \psline(0.7,0)(0.7,0.55)
  \psline(0.7,0.55) (0.25,1)
  \psline{->}(0.7,0.55)(0.3,0.95)
  \psline(0.7,0.55) (1.15,1)
  \psline{->}(0.7,0.55)(1.1,0.95)
  \rput[bl]{0}(0.38,0.2){$c$}
  \rput[br]{0}(1.4,0.8){$b$}
  \rput[bl]{0}(0,0.8){$a$}
  \psline(0.7,0) (0.25,-0.45)
  \psline{-<}(0.7,0)(0.35,-0.35)
  \psline(0.7,0) (1.15,-0.45)
  \psline{-<}(0.7,0)(1.05,-0.35)
  \rput[br]{0}(1.45,-0.4){$b$}
  \rput[bl]{0}(-0.05,-0.4){$a$}
  \endpspicture
.
\end{equation}
This diagram should not be interpreted as worldlines representing motion of the anyons; it signifies only that the combined charge of the two anyons is measured, which, as noted above, does not necessarily involve moving them. When performing topological charge measurements, one must be careful to avoid carrying them out in a manner or configuration that results in undesired effects on the anyonic charge correlations of the system, such as the introduction of unintentional charge entanglement or decoherence of charge entanglement that encodes relevant information.

%
We represent the state encoded in the non-local internal degrees of freedom of a collection of anyons by
\begin{equation}
\left| \psi\left(a,...\right) \right\rangle =
 \pspicture[0.4615](-.4,-.05)(.4,1.25)
  \small
  \psframe[linewidth=0.9pt,linecolor=black,border=0](-0.25,0)(0.25,0.5)
  \rput[bl]{0}(-0.1,0.1){$\psi$}
  \psset{linewidth=0.9pt,linecolor=black,arrowscale=1.5,arrowinset=0.15}
  \psline(0,0.5)(0,1)
  \psline{->}(0,0.5)(0,0.9)
  \rput[bl](-0.1,1.1){$a$}
 \endpspicture
\end{equation}
where we are presently only interested in manipulating the anyon of definite charge $a$, and so leave implicit the additional anyon charge lines emanating from the box, corresponding to the ``$\ldots$'' (normalization factors are absorbed into the box).

In order to teleport the state information encoded by an anyon of definite charge $a$ to another anyon of definite charge $a$, we introduce a particle-antiparticle pair produced from vacuum, given by the state
\begin{equation}
\label{eq:aabarpair}
\left| a,\bar{a};0 \right\rangle = \frac{1}{\sqrt{d_{a}}}
  \pspicture[0.353](0.4,-0.05)(1.6,0.8)
  \small
  \psset{linewidth=0.9pt,linecolor=black,arrowscale=1.5,arrowinset=0.15}
  \psline(1,0)(0.6,0.5)
  \psline(1,0)(1.4,0.5)
   \psline{->}(1,0)(0.7,0.375)
   \psline{->}(1,0)(1.3,0.375)
   \rput[bl]{0}(0.5,0.6){$a$}
   \rput[bl]{0}(1.35,0.6){$\bar{a}$}
  \endpspicture
,
\end{equation}
where $\bar{a}$ is the charge conjugate of $a$, and $0$ is the trivial ``vacuum'' charge, which is interchangeably represented diagrammatically by either a line labeled $0$, a dotted line, or no line at all. The state in Eq.~(\ref{eq:aabarpair}) has maximal anyonic entanglement between its two anyons, and is the analog of the maximally entangled Bell states typically used as the entanglement resource in quantum state teleportation of conventional qubits. Then, for the combined state
\begin{equation}
\label{eq:psiaright}
\left| a,\bar{a};0 \right\rangle_{12} \left| 0,a; \psi\left(a,...\right) \right\rangle_{\left(12\right)3} =
\frac{1}{\sqrt{d_{a}}}
  \pspicture[0.4167](0,-0.6)(2.0,1.8)
  \small
  \psset{linewidth=0.9pt,linecolor=black,arrowscale=1.5,arrowinset=0.15}
  \psline(1,0.5)(1,0)
  \psline(0.2,1.5)(0.6,1)
  \psline(1.8,1.5) (1,0.5)
  \psline(0.6,1) (1,1.5)
   \psline{->}(0.6,1)(0.3,1.375)
   \psline{->}(0.6,1)(0.9,1.375)
   \psline{->}(1,0.5)(1.7,1.375)
   \psline{->}(1,0)(1,0.375)
   \rput[bl]{0}(0.05,1.6){$a$}
   \rput[bl]{0}(0.95,1.6){$\bar{a}$}
   \rput[bl]{0}(1.75,1.6){${a}$}
   \rput[bl]{0}(0.5,0.5){$0$}
   \rput[bl]{0}(1.15,0.2){$a$}
  \psframe[linewidth=0.9pt,linecolor=black,border=0](0.75,-0.5)(1.25,0)
  \rput[bl]{0}(0.9,-0.4){$\psi$}
  \psset{linewidth=0.9pt,linecolor=black,linestyle=dotted,arrowscale=1.5,arrowinset=0.15}
  \psline(0.6,1)(1,0.5)
 \endpspicture
,
\end{equation}
what we would like to do is perform a measurement of the collective charge of anyons $2$ and $3$ for which the result is vacuum charge $0$. This applies the projector $\Pi_{0}^{\left(23\right)}$ and, after applying an isotopy and re-normalizing the state, results in the post-measurement state
\begin{eqnarray}
&& \!\!\!\!\!\!\!\!\!\!\!\!\!\!\!\! d_{a} \Pi_{0}^{\left( 23 \right)} \left| a,\bar{a};0 \right\rangle_{12} \left| 0,a; \psi\left(a,...\right) \right\rangle_{\left(12\right)3}
\notag \\
&&=\frac{1}{\sqrt{d_{a}}}
  \pspicture[0.4167](0.4,-0.6)(1.6,1.8)
  \small
  \psset{linewidth=0.5pt,linecolor=black,arrowscale=1.0,arrowinset=0.15}
  \psline(1,0.25)(1,0)
  \psline(0.6,0.75)(0.8,0.5)
  \psline(1.4,0.75) (1,0.25)
  \psline(0.8,0.5) (1,0.75)
   \psline{->}(1,0)(1,0.1875)
  \psline(0.6,0.75)(0.6,1.5)
  \psline(1,0.75)(1.2,1)
  \psline(1.4,0.75)(1.2,1)
  \psline(1.2,1.25)(1,1.5)
  \psline(1.2,1.25)(1.4,1.5)
  \psline{->}(0.6,1.25)(0.6,1.4375)
  \psline{->}(1.2,1.25)(1.05,1.4375)
  \psline{->}(1.2,1.25)(1.35,1.4375)
   \rput[bl]{0}(0.9,-0.4){$\psi$}
   \scriptsize
   \rput[bl]{0}(0.5,1.6){$a$}
   \rput[bl]{0}(0.95,1.6){$\bar{a}$}
   \rput[bl]{0}(1.35,1.6){${a}$}
   \rput[bl]{0}(1.15,0.1){$a$}
  \psframe[linewidth=0.9pt,linecolor=black,border=0](0.75,-0.5)(1.25,0)
 \endpspicture
= \frac{\varkappa_{a}}{\sqrt{d_{a}}}
  \pspicture[0.4167](0,-0.6)(2.0,1.8)
  \small
  \psset{linewidth=0.9pt,linecolor=black,arrowscale=1.5,arrowinset=0.15}
  \psline(1,0.5)(1,0)
  \psline(0.2,1.5)(1,0.5)
  \psline(1.8,1.5) (1.4,1)
  \psline(1.4,1) (1,1.5)
   \psline{->}(0.6,1)(0.3,1.375)
   \psline{->}(1.4,1)(1.1,1.375)
   \psline{->}(1.4,1)(1.7,1.375)
   \psline{->}(1,0)(1,0.375)
   \rput[bl]{0}(0.05,1.6){$a$}
   \rput[bl]{0}(0.95,1.6){$\bar{a}$}
   \rput[bl]{0}(1.75,1.6){${a}$}
   \rput[bl]{0}(1.25,0.45){$0$}
   \rput[bl]{0}(.7,0.2){$a$}
  \psframe[linewidth=0.9pt,linecolor=black,border=0](0.75,-0.5)(1.25,0)
  \rput[bl]{0}(0.9,-0.4){$\psi$}
  \psset{linewidth=0.9pt,linecolor=black,linestyle=dotted,arrowscale=1.5,arrowinset=0.15}
  \psline(1.4,1)(1,0.5)
 \endpspicture
\notag \\
\label{eq:psialeft}
&&= \varkappa_{a} \left| \bar{a},a;0 \right\rangle_{23} \left| a,0; \psi\left(a,...\right) \right\rangle_{1\left(23\right)}
\end{eqnarray}
where $d_{a}$ in the first line is the state re-normalizing factor [as in Eq.~(\ref{eq:postmeasstate})] and $\varkappa_{a} \equiv d_{a} \left[F_{a}^{a\bar{a}a}\right]_{00}$ is a phase~\footnote{$\varkappa_{a}$ is a gauge-dependent quantity unless $a$ is self-dual, in which case $\varkappa_{a}=\pm 1$ is known as the Frobenius-Schur indicator.}. The transformation from the state in Eq.~(\ref{eq:psiaright}) to the state in Eq.~(\ref{eq:psialeft}) is an anyonic analog of quantum state teleportation that exhibits a path-like behavior specified by the measurement. The encoded state information $\psi$ originally associated with anyon $3$ ends up being associated with anyon $1$ instead, while anyons $2$ and $3$ end up as the maximally entangled pair.

Of course, since measurement outcomes are probabilistic, we cannot guarantee that any given measurement will have the desired outcome. Transformations between the two fusion bases of the three anyons are realized by the unitary $F$-moves
\begin{equation}
  \pspicture[0.4545](0,-0.4)(1.8,1.8)
  \small
  \psset{linewidth=0.9pt,linecolor=black,arrowscale=1.5,arrowinset=0.15}
  \psline(1,0.5)(1,0)
  \psline(0.2,1.5)(1,0.5)
  \psline(1.8,1.5) (1,0.5)
  \psline(0.6,1) (1,1.5)
   \psline{->}(0.6,1)(0.3,1.375)
   \psline{->}(0.6,1)(0.9,1.375)
   \psline{->}(1,0.5)(1.7,1.375)
   \psline{->}(1,0.5)(0.7,0.875)
   \psline{->}(1,0)(1,0.375)
   \rput[bl]{0}(0.05,1.6){$a$}
   \rput[bl]{0}(0.95,1.6){$\bar{a}$}
   \rput[bl]{0}(1.75,1.6){${a}$}
   \rput[bl]{0}(0.5,0.5){$e$}
   \rput[bl]{0}(.95,-0.3){$a$}
  \endpspicture
= \sum_{f} \left[F_{a}^{a\bar{a}a}\right]_{ef}
 \pspicture[0.4545](0,-0.4)(1.8,1.8)
  \small
  \psset{linewidth=0.9pt,linecolor=black,arrowscale=1.5,arrowinset=0.15}
  \psline(1,0.5)(1,0)
  \psline(0.2,1.5)(1,0.5)
  \psline(1.8,1.5) (1,0.5)
  \psline(1.4,1) (1,1.5)
   \psline{->}(0.6,1)(0.3,1.375)
   \psline{->}(1.4,1)(1.1,1.375)
   \psline{->}(1,0.5)(1.7,1.375)
   \psline{->}(1,0.5)(1.3,0.875)
   \psline{->}(1,0)(1,0.375)
   \rput[bl]{0}(0.05,1.6){$a$}
   \rput[bl]{0}(0.95,1.6){$\bar{a}$}
   \rput[bl]{0}(1.75,1.6){${a}$}
   \rput[bl]{0}(1.25,0.45){$f$}
   \rput[bl]{0}(.95,-0.3){$a$}
  \endpspicture
\end{equation}
and their inverses. This indicates that the probability of collective charge measurement outcome $f$ for anyons $2$ and $3$, given that anyons $1$ and $2$ were in the definite charge $e$ fusion channel, and vice-versa, is
\begin{equation}
\text{Prob}\left(e|f \right) =\text{Prob}\left(f|e \right)=\left| \left[ F^{a \bar{a} a}_{a} \right]_{ef} \right|^{2} .
\end{equation}

But we see that we can, in a sense, undo an undesired measurement outcome of anyons $2$ and $3$ ($f \neq 0$), by subsequently performing a measurement of anyons $1$ and $2$, as long as the measurement processes are non-demolitional. This returns the system to a state in which anyons $1$ and $2$ have definite collective charge $e$ (not necessarily $0$ now), but otherwise leaves the encoded state information undisturbed. We may now perform a measurement of anyons $2$ and $3$ again, with an entirely new chance of obtaining the desired outcome ($f=0$). This procedure may be repeated until we obtain the desired measurement outcome, obtaining a string of measurement outcomes $M=\left\{ e_{1},f_{1},\ldots,e_{n},f_{n} \right\}$ (including the initialization $e_{1}$ for later convenience in representing this process), where $e_{1}=f_{n}=0$, and so we call it ``forced measurement.'' The probability of obtaining the desired outcome at the $j^{th}$ measurement attempt in this procedure is
\begin{equation}
\text{Prob}\left(f_{j}=0|e_{j}\right) = \left| \left[ F^{a \bar{a} a}_{a} \right]_{e_{j}0} \right|^{2}
=d_{e_{j}}/d_{a}^{2} \geq d_{a}^{-2}
,
\end{equation}
where $d_{x}$ is the quantum dimension of the anyonic charge $x$, and $d_{x} \geq 1$ with equality iff $x$ is Abelian (e.g. $d_{0}=1$). The average number of attempts until a desired outcome is achieved in a forced measurement is thus $\left\langle n \right\rangle \leq d_{a}^{2}$, and the probability of needing $n>N$ attempts to obtain the desired outcome is $\text{Prob}\left(f_{1},\ldots,f_{N} \neq 0 \right)\leq \left( 1-d_{a}^{-2} \right)^{N}$, i.e. failure is exponentially suppressed in the number of attempts.

Forced measurement is a probabilistically determined adaptive series of measurements. More precisely, the measurements to be carried out are pre-determined, but the number of times $n$ that they need to be carried out is probabilistically determined based on the first attainment of the desired outcome $f_{n}=0$. The resulting operator representing such a forced measurement transformation acting on an $e_{1}=0$ initialized state, with the state re-normalizing factor $A$ included, is
\begin{eqnarray}
&& \!\!\!\!\!\!\!\! \breve{\Pi}_{M}^{\left(23\leftarrow 12 \right)}
=\frac{1}{A} \Pi_{f_{n}=0}^{\left(23 \right)} \Pi_{e_{n}}^{\left(12 \right)}
\ldots \Pi_{f_{1}}^{\left(23 \right)} \Pi_{e_{1}=0}^{\left(12 \right)}
\notag \\
&&
= \frac{\sqrt{ d_{f_{1}} \ldots d_{e_{n}} } }{A d_{a}^{2n}}
 \pspicture[0.4787](-0.8,-0.4)(0.7,4.3)
  \psset{linewidth=0.5pt,linecolor=black,arrowscale=1.0,arrowinset=0.15}
  \psline(-0.4,0)(-0.2,0.25)
  \psline(0,0)(-.2,0.25)
  \psline(-0.4,0.75)(-0.2,0.5)
  \psline(0,0.75)(-.2,0.5)
  \psline(0.4,0)(0.4,0.75)
  \psline{-<}(-0.2,0.25)(-0.35,0.0625)
  \psline{-<}(-0.2,0.25)(-0.05,0.0625)
  \psline{-<}(0.4,0.25)(0.4,0.0625)
  \psline(-0.4,0.75)(-0.4,1.5)
  \psline(0,0.75)(0.2,1)
  \psline(0.4,0.75)(0.2,1)
  \psline(0.2,1)(0.2,1.25)
  \psline(0.2,1.25)(0,1.5)
  \psline(0.2,1.25)(0.4,1.5)
  \psline{->}(0.2,1)(0.2,1.1875)
  \psline(-0.4,2.5)(-0.2,2.75)
  \psline(0,2.5)(-0.2,2.75)
  \psline(-0.4,3.25)(-0.2,3)
  \psline(0,3.25)(-0.2,3)
  \psline(-0.2,2.75)(-0.2,3)
  \psline(0.4,2.5)(0.4,3.25)
  \psline{->}(-0.2,2.75)(-0.2,2.9375)
  \psline(-0.4,3.25)(-0.4,4)
  \psline(0,3.25)(0.2,3.5)
  \psline(0.4,3.25)(0.2,3.5)
  \psline(0.2,3.75)(0,4)
  \psline(0.2,3.75)(0.4,4)
  \psline{->}(-0.4,3.75)(-0.4,3.9375)
  \psline{->}(0.2,3.75)(0.05,3.9375)
  \psline{->}(0.2,3.75)(0.35,3.9375)
   \psset{linewidth=2pt,linecolor=black,linestyle=dotted,dotsep=5pt}
  \psline(0,1.75)(0,2.25)
   \psset{linewidth=0.25pt,linecolor=black,linestyle=dashed}
  \psline(-0.6,0.75)(0.6,0.75)
  \psline(-0.6,1.5)(0.6,1.5)
  \psline(-0.6,2.5)(0.6,2.5)
  \psline(-0.6,3.25)(0.6,3.25)
   \scriptsize
   \rput[bl]{0}(-0.5,4.1){$a$}
   \rput[bl]{0}(-0.05,4.1){$\bar{a}$}
   \rput[bl]{0}(0.35,4.1){${a}$}
   \rput[bl]{0}(-0.5,-0.2){$a$}
   \rput[bl]{0}(-0.05,-0.2){$\bar{a}$}
   \rput[bl]{0}(0.35,-0.2){${a}$}
   \rput[bl]{0}(0.35,1){${f_{1}}$}
   \rput[bl]{0}(-0.6,2.8){${e_{n}}$}
 \endpspicture
= \frac{e^{i \varphi_{ M}} }{d_{a}}
 \pspicture[0.4627](0,-1.45)(2,1.9)
  \small
  \psset{linewidth=0.9pt,linecolor=black,arrowscale=1.5,arrowinset=0.15}
  \psline(1,0)(1.8,-1)
  \psline(0.6,-0.5)(1,-1)
  \psline(0.6,-0.5)(0.2,-1)
  \psline(1,0.5)(1,0)
  \psline(0.2,1.5)(1,0.5)
  \psline(1.8,1.5)(1.4,1)
  \psline(1,1.5)(1.4,1)
   \psline{->}(0.6,1)(0.3,1.375)
   \psline{->}(1.4,1)(1.1,1.375)
   \psline{->}(1.4,1)(1.7,1.375)
   \psline{-<}(0.6,-0.5)(0.3,-0.875)
   \psline{-<}(0.6,-0.5)(0.9,-0.875)
   \psline{-<}(1.4,-0.5)(1.7,-0.875)
   \rput[bl]{0}(0.05,1.6){$a$}
   \rput[bl]{0}(0.95,1.6){$\bar{a}$}
   \rput[bl]{0}(1.75,1.6){${a}$}
   \rput[bl]{0}(0.05,-1.3){$a$}
   \rput[bl]{0}(0.95,-1.3){$\bar{a}$}
   \rput[bl]{0}(1.75,-1.3){${a}$}
  \endpspicture
\notag \\
&&
\!\!\!\!\!\!\!\!\!\!\!\! = e^{i \varphi_{ M}}\left| \bar{a},a;0 \right\rangle_{23} \left| a,0;a \right\rangle_{1\left(23\right)}
\left\langle 0,a;a \right|_{\left(12\right)3} \left\langle a,\bar{a};0 \right|_{12}
\end{eqnarray}
(For improved clarity, we inserted dashed lines to partition the diagram into sections corresponding to each individual topological charge measurement projector.) The diagrammatic evaluation in the second line is made simple by applying charge conservation to the diagrammatic web in the middle of the left hand side. It may be treated as a blob with only one charge $a$ line entering it and one charge $a$ line leaving it, and so can simply be replaced by a charge $a$ line running straight through (as in the diagram on right hand side) multiplied by a constant factor. This constant cancels with the other factors to leave the appropriate normalization and an overall phase $e^{i \varphi_{ M}}$, which depends on the measurement outcomes $M$. Thus, through forced measurement we obtain the transformation from Eq.~(\ref{eq:psiaright}) to Eq.~(\ref{eq:psialeft})
\begin{eqnarray}
&&\breve{\Pi}_{M}^{\left(23\leftarrow 12 \right)} \left| a,\bar{a};0 \right\rangle_{12} \left| 0,a; \psi\left(a,...\right) \right\rangle_{\left(12\right)3} \notag \\
&& \qquad \quad =e^{i \varphi_{M}} \left| \bar{a},a;0 \right\rangle_{23} \left| a,0; \psi\left(a,...\right) \right\rangle_{1\left(23\right)}
\end{eqnarray}
giving us anyonic teleportation.~\footnote{A forced measurement procedure may similarly be applied to state teleportation of conventional qubits, though this is typically not emphasized, since in such cases it is easier to instead correct for ``undesired'' outcomes by applying a Pauli operator.}

\begin{figure}[t!]
\begin{center}
  \includegraphics[scale=0.2]{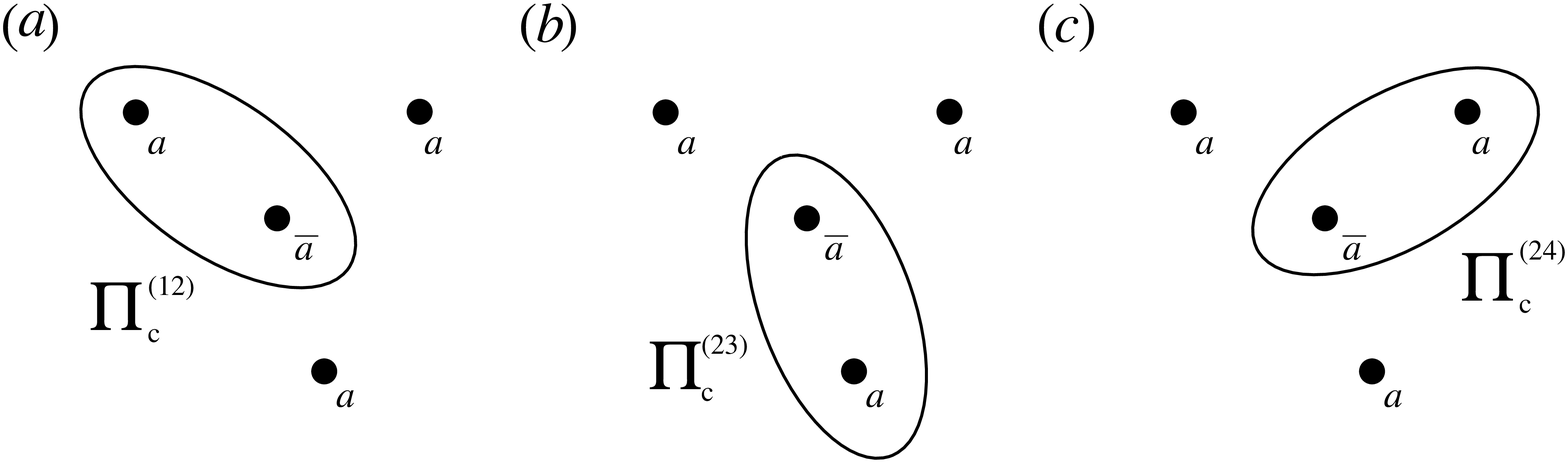}
  \caption{Projective topological charge measurements of pairs of anyons (a) $1$ and $2$, (b) $2$ and $3$, and (c) $2$ and $4$ are used to implement ``forced measurement'' anyonic state teleportation, which is used to produce braiding transformations as in Eqs.~(\ref{eq:R},\ref{eq:R*}).}
  \label{fig:measurements}
\end{center}
\end{figure}

Now we can produce the braiding transformations for two anyons of definite charges $a$
\begin{equation}
R_{aa}=
 \pspicture[0.4](-0.1,0)(1.4,1)
  \psset{linewidth=0.9pt,linecolor=black,arrowscale=1.5,arrowinset=0.15}
  \psline(0.96,0.05)(0.2,1)
  \psline{->}(0.96,0.05)(0.28,0.9)
  \psline(0.24,0.05)(1,1)
  \psline[border=2pt]{->}(0.24,0.05)(0.92,0.9)
  \rput[bl]{0}(-0.05,0.1){$a$}
  \rput[br]{0}(1.25,0.1){$a$}
  \endpspicture
,\qquad
R_{aa}^{-1}= R_{aa}^{\dag}=
 \pspicture[0.4](-0.1,0)(1.4,1)
  \psset{linewidth=0.9pt,linecolor=black,arrowscale=1.5,arrowinset=0.15}
  \psline{->}(0.24,0.05)(0.92,0.9)
  \psline(0.24,0.05)(1,1)
  \psline(0.96,0.05)(0.2,1)
  \psline[border=2pt]{->}(0.96,0.05)(0.28,0.9)
  \rput[bl]{0}(-0.05,0.1){$a$}
  \rput[br]{0}(1.25,0.1){$a$}
  \endpspicture
,
\end{equation}
by introducing a maximally entangled anyon pair and performing three forced measurements
\begin{eqnarray}
&& \!\!\!\! \breve{\Pi}_{M_{3}}^{\left(23\leftarrow 24 \right)} \breve{\Pi}_{M_{2}}^{\left(24\leftarrow 12 \right)} \breve{\Pi}_{M_{1}}^{\left(12\leftarrow 23 \right)}
\notag \\
&&= \frac{e^{i \varphi_{M}}}{d_{a}^{3}}
 \pspicture[0.5](-1,-0.4)(1,2.5)
  \psset{linewidth=0.5pt,linecolor=black,arrowscale=1.0,arrowinset=0.15}
  \psline(-0.6,0)(0.2,0.75)
  \psline(-0.2,0)(0,0.25)
  \psline(0.2,0)(0,0.25)
  \psline(0.6,0)(0.6,0.75)
  \psline(-0.4,0.5)(-0.2,0.75)
  \psline(-0.4,0.5)(-0.6,0.75)
  \psline{-<}(0,0.25)(-0.15,0.0625)
  \psline{-<}(0,0.25)(0.15,0.0625)
  \psline{-<}(0.6,0.25)(0.6,0.0625)
  \psline{-<}(0.2,0.75)(-0.5333333333,0.0625)
  \psline{->}(-0.4,0.5)(-0.55,.6875)
  \psline(-0.6,0.75)(-0.4,1)
  \psline(-0.2,0.75)(-0.4,1)
  \psline(0.6,0.75)(-0.6,1.5)
  \psline(0.2,1.25)(-0.2,1.5)
  \psline(0.2,1.25)(0.6,1.5)
  \psline{->}(0.2,1.25)(0.5,1.4375)
  \psline[border=1.5pt](0.2,0.75)(0.2,1.5)
  \psline(-0.6,1.5)(-0.6,2.25)
  \psline(-0.2,1.5)(0.2,1.75)
  \psline(0.6,1.5)(0.2,1.75)
  \psline(0,2)(-0.2,2.25)
  \psline(0,2)(0.2,2.25)
  \psline[border=1.5pt](0.2,1.5)(0.6,2.25)
  \psline{->}(-0.6,1.5)(-0.6,2.1875)
  \psline{->}(0,2)(-0.15,2.1875)
  \psline{->}(0,2)(0.15,2.1875)
  \psline{->}(0.2,1.5)(0.5666666666,2.1875)
   \psset{linewidth=0.25pt,linecolor=black,linestyle=dashed}
  \psline(-0.8,0.75)(0.8,0.75)
  \psline(-0.8,1.5)(0.8,1.5)
   \scriptsize
   \rput[bl]{0}(-0.7,2.35){$a$}
   \rput[bl]{0}(-0.3,2.35){$\bar{a}$}
   \rput[bl]{0}(0.15,2.35){$a$}
   \rput[bl]{0}(0.55,2.35){${a}$}
   \rput[bl]{0}(-0.7,.45){$a$}
   \rput[bl]{0}(0.5,1.25){$a$}
   \rput[bl]{0}(-0.7,-0.2){$a$}
   \rput[bl]{0}(-0.3,-0.2){$\bar{a}$}
   \rput[bl]{0}(0.15,-0.2){$a$}
   \rput[bl]{0}(0.55,-0.2){${a}$}
 \endpspicture
= \frac{e^{i \varphi_{M}}}{d_{a}}
 \pspicture[0.5185](-1.5,-0.4)(1.8,2.3)
  \psset{linewidth=0.9pt,linecolor=black,arrowscale=1.5,arrowinset=0.15}
  \psline(-0.4,0)(0,0.5)
  \psline(0.4,0)(0,0.5)
  \psline(1.2,0)(-1.2,2)
  \psline[border=2.5pt](-1.2,0)(1.2,2)
  \psline(0,1.5)(-0.4,2)
  \psline(0,1.5)(0.4,2)
  \psline{-<}(0,0.5)(0.3,0.125)
  \psline{-<}(0,0.5)(-0.3,0.125)
  \psline{->}(0,1.5)(-0.3,1.875)
  \psline{->}(0,1.5)(0.3,1.875)
  \psline{-<}(-1.2,2)(-0.6,1.5)
  \psline{-<}(1.2,2)(0.6,1.5)
   \rput[bl]{0}(-1.3,2.1){$a$}
   \rput[bl]{0}(-0.5,2.1){$\bar{a}$}
   \rput[bl]{0}(0.35,2.1){$a$}
   \rput[bl]{0}(1.15,2.1){$a$}
   \rput[bl]{0}(-1.3,-0.3){$a$}
   \rput[bl]{0}(-0.5,-0.3){$\bar{a}$}
   \rput[bl]{0}(0.35,-0.3){$a$}
   \rput[bl]{0}(1.15,-0.3){$a$}
 \endpspicture
\notag \\
\label{eq:R}
&&= e^{i \varphi_{M}} R_{aa}^{\left(14\right)} \otimes  \left| \bar{a},a;0 \right\rangle_{23} \left\langle \bar{a},a;0 \right|_{23}
,
\end{eqnarray}
where $\varphi_{M} = \varphi_{M_{1}} + \varphi_{M_{2}} + \varphi_{M_{3}}$, and similarly
\begin{eqnarray}
&&\breve{\Pi}_{M_{3}}^{\left(23\leftarrow 12 \right)} \breve{\Pi}_{M_{2}}^{\left(12\leftarrow 24 \right)} \breve{\Pi}_{M_{1}}^{\left(24\leftarrow 23 \right)} \notag \\
\label{eq:R*}
&& \qquad =e^{i \varphi_{M}} \left[ R_{aa}^{\left(14\right)} \right]^{-1} \otimes  \left| \bar{a},a;0 \right\rangle_{23} \left\langle \bar{a},a;0 \right|_{23}
.
\end{eqnarray}
The spatial configuration of these anyons and the measurements used are shown in Fig.~\ref{fig:measurements}. An important point to emphasize is that the entanglement resource (the maximally entangled anyon pair) is fully replenished and returned to its original location at the end of these processes. This allows such measurement-generated braiding transformations to be employed repeatedly, without exhausting the resources.

\begin{figure}[t!]
\begin{center}
  \includegraphics[scale=0.4]{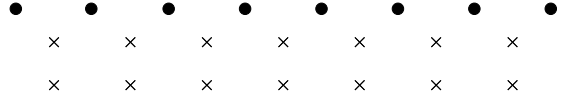}
  \caption{A quasi-one-dimensional array of stationary anyons used for measurement-only topological quantum computing. Maximally entangled pairs of anyons (denoted by X's) are situated between adjacent pairs of computational anyons (denoted by dots) to facilitate measurement induced braiding transformations used to generate computational gates.}
  \label{fig:quasiarray}
\end{center}
\end{figure}

It is now straightforward to apply these results to TQC. We arrange our computational anyons in a linear array and distribute maximally entangled pairs (more or less) between them, forming a quasi-one-dimensional array, as in Fig.~\ref{fig:quasiarray}. These anyons all remain stationary and computational gates on the topological qubits are implemented via measurements. The relations in Eqs.~(\ref{eq:R},\ref{eq:R*}) give a map between braiding transformations (with an irrelevant overall phase) and measurements, so one may simply use the established techniques of generating computational gates from braiding transformations to determine the series of measurements that should be performed to implement a particular quantum algorithm. If the computational anyons have self-dual charge $a=\bar{a}$, we can use a more economical distribution of entanglement resource anyons, situating only one anyon from each maximally entangled pair between each adjacent pair of computational anyons (i.e. the second row of X's in Fig.~\ref{fig:quasiarray} is not needed). For TQC models in which the computational anyons do not all have the same anyonic charge, the same anyonic teleportation principles may be applied, but a greater number of entanglement resource anyons will be needed.

The basic operation in this TQC scheme is a measurement, which is not topological in the sense that what is measured is a non-topological quantity (e.g. current) that infers information regarding the topological charge through an approximate relation dictated by the particular experimental setup used. This seems less robust than the physical braiding operation it is used to replace, however, the measured quantity is the topological charge of a pair of quasiparticles, which is conserved so long as all other quasiparticles are far away. Therefore, there is still a form of topological protection, though a perhaps slightly weaker one than in the braiding model, since measurements are not entirely accurate. In the qubit model of measurement-only quantum computation, on the other hand, there is always the danger of interaction with the environment causing an error in a qubit that we need to measure.

The Ising or $\text{SU}\left(2\right)_{2}$ anyons, such as those that are likely to occur in second Landau level FQH states~\cite{Moore91,Bonderson07d}, have anyonic charges $0, \frac{1}{2},1$ (a.k.a. $I,\sigma,\psi$ respectively). For TQC models based on such systems~\cite{DasSarma05}, the computational anyons should have charge $a=\frac{1}{2}$, and the measurement outcomes will be $e_{j},f_{j}=0,1$. The quantum dimensions of these are $d_{0}=d_{1}=1$ and $d_{\frac{1}{2}}=\sqrt{2}$. Unfortunately, quasiparticle braiding in these anyon models is not quite computationally universal, and so must be supplemented by operations that are topologically unprotected~\cite{Bravyi06} or involve changing the topology of the system~\cite{FNW05a,FNW05b}. Furthermore, the TQC model based on Ising/SU$\left(2\right)_{2}$ is a special case in which the interferometry measurements employed (i.e. that of pairs of charge $\frac{1}{2}$ anyons) are in fact also projective measurements, because the measurement outcome charges $0$ and $1$ are Abelian in these anyon models, satisfying Eq.~(\ref{eq:intisproj}). This allows the methods described in this letter to be applied directly when using interferometry measurements in such models.

Fibonacci anyons, which do have computationally universal braiding and might also occur in FQH states~\cite{Read99}, have anyonic charges $0,1$ (a.k.a. $I,\varepsilon$ respectively). Computational anyons must have charge $a=1$, which has quantum dimension $d_{1}=\phi = \frac{1+\sqrt{5}}{2}$ (the Golden ratio), and the measurement outcomes will be $e_{j},f_{j}=0,1$.

For TQC models based on $\text{SU}\left(2\right)_{k}$ anyons~\cite{Freedman02a,Freedman02b}, which have computationally universal braiding for $k=3$ and all $k \geq 5$, the computational anyons have charge $a=\frac{1}{2}$, which has quantum dimension $d_{\frac{1}{2}}=2\cos\frac{\pi}{k+2}$, and the measurement outcomes will be $e_{j},f_{j}=0,1$.

In contrast with projective measurement, interferometrical measurement of topological charge is not quite as simple and requires a density matrix formulation. In the asymptotic limit, interferometry may effectively be treated as a projective measurement of the target anyons' collective charge, together with decoherence of anyonic charge entanglement between the target anyons and those exterior to the interferometry region~\cite{Bonderson07a,Bonderson07b,Bonderson07c}. Specifically, anyonic charge entanglement encoded in the anyons' density matrix as non-trivial charge lines connecting the interior and exterior interferometry regions will generically decohere due to the interferometrical measurement process, leaving only density matrix components with vacuum or no charge lines connecting these regions. An interferometrical measurement is the same as a projective measurement only when the topological charge measurement outcome $c$ obeys the property
\begin{equation}
\label{eq:intisproj}
 \pspicture[0.5357](-0.4,-0.4)(0.4,1)
  \psset{linewidth=0.5pt,linecolor=black,arrowscale=1.0,arrowinset=0.15}
  \psline(-0.2,0)(0,0.25)
  \psline(0.2,0)(0,0.25)
  \psline(0,0.5)(-0.2,0.75)
  \psline(0,0.5)(0.2,0.75)
  \psline{-<}(0,0.25)(0.15,0.0625)
  \psline{-<}(0,0.25)(-0.15,0.0625)
  \psline{->}(0,0.5)(-0.15,0.6925)
  \psline{->}(0,0.5)(0.15,0.6925)
   \scriptsize
   \rput[bl]{0}(-0.3,0.85){$\bar{c}$}
   \rput[bl]{0}(0.15,0.85){$c$}
   \rput[bl]{0}(-0.3,-0.3){$\bar{c}$}
   \rput[bl]{0}(0.15,-0.3){$c$}
 \endpspicture
=
 \pspicture[0.5357](-0.4,-0.4)(0.4,1)
  \psset{linewidth=0.5pt,linecolor=black,arrowscale=1.0,arrowinset=0.15}
  \psline(-0.2,0)(-0.2,0.75)
  \psline(0.2,0)(0.2,0.75)
  \psline{->}(-0.2,0)(-0.2,0.5)
  \psline{->}(0.2,0)(0.2,0.5)
   \scriptsize
   \rput[bl]{0}(-0.3,0.85){$\bar{c}$}
   \rput[bl]{0}(0.15,0.85){$c$}
   \rput[bl]{0}(-0.3,-0.3){$\bar{c}$}
   \rput[bl]{0}(0.15,-0.3){$c$} \endpspicture
,
\end{equation}
which is the case iff $c$ is Abelian. Because of this important distinction, the forced measurement process described here is not generally applicable for all anyon models when using interferometry for topological charge measurement (though it is in certain special cases, such as Ising/SU$\left(2\right)_{2}$). Fortunately, an interferometrical version of forced measurement may be attained by a slight procedural modification that includes performing additional measurements that involve up to $8$ computational anyons~\cite{Bonderson08b}.

We have described an anyonic analog of quantum state teleportation and showed how it may be utilized to generate the braiding transformations $R$ (up to a phase) from an adaptive series of non-demolitional topological charge measurements, rather than physical movement of the computational anyons. This provides a new, ``measurement-only'' approach to TQC, in which topological charge measurement is the only primitive employed. In particular, we can replace the primitives 2 and 3 from the introduction with:

2$^{\prime}$. Non-demolitional measurement of collective topological charge of anyons, for qubit initialization and readout, and to implement the desired computational gates.

\noindent Exuviating the need for quasiparticle braiding, once considered an essential primitive of TQC, in favor of topological charge measurement, which is unavoidably necessary for qubit readout, may prove to be an essential disencumberment in the implementation of TQC.


\end{document}